\documentclass{ws-ijmpa}

\usepackage[super,compress]{cite}
\usepackage{graphicx, amssymb, latexsym, collref, epsfig, amsmath, bbm, color, hyperref}
\hypersetup{colorlinks,urlcolor=black,citecolor=black,linkcolor=black,filecolor=black}
\usepackage{breakurl}
\usepackage[utf8]{inputenc} 

\def\tr{{\rm Tr\,}}

\def\bea{\begin{eqnarray}}
\def\eea{\end{eqnarray}}
\def\nn{\nonumber}

\def\lmatrix{\left(\begin{array}}
\def\rmatrix{\end{array}\right)}
\newcommand{\RS}{r}

\begin{document}
\markboth{Daniel Nogradi and Agostino Patella}{Strong dynamics}

%
\catchline{}{}{}{}{}
%

\title{Strong dynamics, composite Higgs and the conformal window}

\author{Daniel Nogradi}

\address{Eotvos University, Institute for Theoretical Physics, Budapest 1117, Hungary \\
MTA-ELTE Lendulet Lattice Gauge Theory Research Group, Budapest 1117, Hungary}

\author{Agostino Patella}

\address{CERN-TH, CH-1211 Geneva, Switzerland \\
Centre for Mathematical Sciences, Plymouth University, Plymouth, PL4 8AA, UK}

\maketitle

\begin{abstract}
We review recent progress in the lattice investigations of near-conformal non-abelian gauge theories relevant for
dynamical symmetry breaking and model building of composite Higgs models.
The emphasis is placed on the mass spectrum and the running renormalized coupling.
The role of a light composite scalar isosinglet particle as a composite Higgs particle is highlighted.

\vspace{5mm}

\begin{center}
May 2, 2016
\end{center}

\end{abstract}

%
\catchline{}{}{}{}{}
%

\tableofcontents

\newpage

\section{Introduction}

Even though the Standard Model and its electroweak sector in particular are extraordinarily successful in terms of both
experimental and theoretical precision the idea of dynamical symmetry breaking came about already in the late 70's
\cite{Weinberg:1979bn, Susskind:1978ms, Dimopoulos:1979es}. The main
motivation was and continues to be naturalness and the associated fine tuning problem. In the
early technicolor paradigm, scaled up QCD with $\Lambda \sim O(TeV)$ was envisioned to take the place of the Higgs sector,
and spontaneous chiral symmetry breaking would be responsible for electroweak
symmetry breaking. The resulting Goldstone bosons or techni pions would be 
eaten by the $W$ and $Z$ bosons and hence the latter would become massive.
In particular the model would be either Higgsless or would feature a heavy composite Higgs, analogous to the $\sigma$ or
$f_0$ meson of QCD, at least according to early expectations.

The initial
proposal faced numerous problems including a potentially large $S$-parameter \cite{Peskin:1991sw} 
and the tension between the observed fermion
masses and potentially large flavor changing neutral currents \cite{Appelquist:1986an}. 
The idea of walking \cite{Holdom:1984sk,Appelquist:1987fc}
was introduced to circumvent some of these issues by
assuming that the renormalized coupling was running slowly between two well separated energy scales $\Lambda_{TC}$
and $\Lambda_{ETC}$, where $ETC$ stands for extended technicolor\cite{Dimopoulos:1979es,Eichten:1979ah}. 
In addition a large mass anomalous dimension was assumed to be generated
along the renormalization group trajectory. The large anomalous dimension would guarantee that flavor changing
neutral currents remain small while the mass of the top quark is the correct one. 
At the same time the precise mechanism for fermion mass generation, dubbed extended technicolor, 
is pushed to a high scale
$\Lambda_{ETC}$ and essentially decouples from the mechanism of electroweak symmetry breaking.
Hence the techni gauge sector responsible for electroweak symmetry breaking is thought to be an effective theory only,
even though in principle it could be a fundamental theory as QCD.

A natural way to look for models with a coupling that walks is by considering non-abelian gauge theories in the
parameter space $(G,N_f,R)$ where $G$ is the gauge group 
and $N_f$ the number of massless fermion flavors in representation $R$. In
the original technicolor proposals, usually $G=SU(N)$ and the fundamental representation was considered. More generally, once
$G$ and $R$ are fixed, $N_f$ may be viewed as a variable and the model may be in one of three phases depending on the value of $N_f$. Clearly, if $N_f$ is too high asymptotic freedom is lost because the first $\beta$-function coefficient
will cease to be negative and the theory is trivial. 
The requirement of asymptotic freedom limits $N_f < N_f^{AF}$ from above and the bound $N_f^{AF}$
is obtained exactly by the 1-loop $\beta$-function.
Just below this upper bound the model has a Banks-Zaks fixed point
with a coupling that is small and can be obtained from a 2-loop calculation \cite{Caswell:1974gg,Banks:1981nn}.
Consequently such a model is a weakly coupled
conformal field theory at long distances and all of its properties such as anomalous dimensions, etc., are calculable
perturbatively in a reliable way. As $N_f$ is decreased further the fixed point coupling grows. 
At some critical value $N_f^*$ the coupling becomes strong enough to generate spontaneous 
symmetry breaking and a dynamical scale like in QCD.
Further decreasing $N_f$
towards zero does not change the infrared dynamics in a substantial way although as we will see the detailed properties
will be very sensitive to the difference $N_f^* - N_f$. The range $N_f^* < N_f < N_f^{AF}$ is called the conformal
window and of course depends on the gauge group $G$ and the representation $R$. In contrast to the upper end of
the conformal window $N_f^{AF}$, the lower end of the conformal window $N_f^*$ is not calculable in perturbation theory. 

It should be noted that the above picture assumes that the flavor number can change continuously which is obviously not
the case. For fixed $G$ and $R$ there is only a discrete set of flavor numbers below the upper end of the conformal
window $N_f^{AF}$ and the arguments based on a continuous change in $N_f$ may or may not be a good guide. This
state of affairs also calls for non-perturbative lattice calculations which in principle can scan all available flavor
numbers $N_f < N_f^{AF}$ and determine the infrared properties for each. 

A relatively recent development was the realization that 
higher dimensional representations $R$ have a lower $N_f^*$ and
hence lower fermion flavor number would be needed for the theory just below the conformal window. As a result the
$S$-parameter can be hoped to be lower, relative to the fundamental representation, and potentially consistent with electroweak
precision data \cite{Sannino:2004qp, Hong:2004td, Dietrich:2005jn}. 
Compatibility of LHC data and a composite Higgs of the type considered here,
including its couplings to the $W$ and $Z$ gauge bosons was scrutinized recently in detail \cite{Belyaev:2013ida}.

The present review has a very limited scope and focuses on a selection of topics mostly related to lattice studies. The
literature on the subject of dynamical electroweak symmetry breaking, technicolor in its many variants and IR-conformal
gauge theories is vast. Extensive reviews on the phenomenological, experimental and formal aspects are available
\cite{Chivukula:2000mb,Hill:2002ap,Lane:2002wv,Martin:2008cd,Sannino:2009za,Piai:2010ma}
as well as more extended reviews of the lattice aspects \cite{DeGrand:2009mt, DelDebbio:2014ata, DeGrand:2015zxa}. 

In section \ref{strong} we discuss the possibility of a light scalar in strongly coupled
gauge theories. In section \ref{conformaland} a pedagogical and elementary introduction to the main differences between chirally
broken and conformal gauge theories is given, focusing on the scaling properties of the mass spectrum. 
In section \ref{latticeaspects} we discuss
some lattice specific issues and we review the main results on the spectrum for a number of models. In section
\ref{sec:running} we discuss different definitions of the running coupling, and review related lattice results. 
Finally in section \ref{outlook} we end with an outlook.

It should be noted that due to the lack of space various very useful approaches of distinguishing conformal and
chiral symmetry broken models and studying their properties on the lattice are not discussed in the present review.
These include finite temperature studies
\cite{Deuzeman:2008sc,Kogut:2010cz,Kogut:2011bd,Kogut:2011ty,Kogut:2014kla,Kogut:2015zta}, finite size scaling
\cite{DeGrand:2009hu ,DeGrand:2011cu, Fodor:2012et, Cheng:2013xha, Lombardo:2014pda, Athenodorou:2014eua, Lin:2015zpa}, radial quantization
\cite{Brower:2012vg, Neuberger:2014pya}, non-degenerate
fermion masses for many flavors in order to interpolate between different flavor numbers 
\cite{Brower:2014dfa, Brower:2015owo} and the spectral properties of the Dirac operator
\cite{Cheng:2011ic,Patella:2012da,Cheng:2013eu,Fodor:2014zca,Perez:2015yna}.

\section{Strong dynamics and a light scalar}
\label{strong}

Even though the original technicolor paradigm of the late 70's envisioned a Higgsless electroweak sector or one with a
heavy Higgs, the possibility of a light composite Higgs was nevertheless actively debated \cite{Yamawaki:1985zg, Holdom:1987yu}.
It is important to note that the scalar isosinglet mass, naturally, needs to be measured
against some other mass scale and its lightness will depend on what scale it is compared to. 
From a phenomenological point of
view the relevant comparison is the mass ratio of the scalar, $m_{\sigma}$, and some other massive state (for instance the
vector isotriplet meson $m_\varrho$) which also stays non-zero in the
chiral limit, assuming the model breaks chiral symmetry. 
This ratio would indicate how far the light scalar is separated
from the tower of other massive particle states. 

Recent lattice simulations in the $(G,N_f,R)$ parameter space of non-abelian gauge theories show that as
the model approaches the conformal window from below the scalar isosinglet meson 
in fact becomes light, relative to the vector meson, $\varrho$ in QCD.
The lattice evidence comes primarily from simulations of $SU(3)$ gauge theory. 
In the $N_f = 8$ fundamental model with $SU(3)$ lattice calculations
indicate that approximately $m_\sigma / m_\varrho \sim 1/2$ can be reached with the available lattice volumes and fermion masses 
\cite{Aoki:2014oha, Rinaldi:2015axa, Appelquist:2016viq}. Another
model which seems to be close to the conformal window, $SU(3)$ with $N_f = 2$ sextet fermions, also features a light
scalar according to lattice calculations. In this model approximately $m_\sigma / m_\varrho \sim 1/4$ was observed 
\cite{Fodor:2014pqa, Fodor:2015eea, Fodor:2015vwa} predicting an even larger separation between the scalar and the rest
of the spectrum. These observations make it plausible that a composite Higgs may emerge from a near-conformal gauge
theory with its $125\; GeV$ mass obtained after electro-weak corrections are taken into account, most notably the
contribution of the top quark \cite{Foadi:2012bb}.

The generation of the Standard Model fermion masses is still left to higher scales and the 
models are still thought of as effective theories only.

The natural question is what mechanism produces a light scalar out of a strongly interacting non-abelian
gauge theory. Again it is important to note what we mean by light. Since just below the conformal window chiral symmetry
is broken, all states have masses $\sim \Lambda$ except the Goldstones. 
What is required is that the ratio of the scalar mass and all other massive states is small. Clearly, there is no small
parameter in the theory for fixed $(G,N_f,R)$. One could think of $N_f^* - N_f$ as a small parameter, if
one approaches the conformal window from below (leaving aside the issue that $N_f$ is discrete). Then one would be
tempted to further argue that as $N_f^* - N_f$ goes to zero, the theory becomes conformal and the $\beta$-function vanishes.
Hence, as this line of argumentation would go, the mass of the scalar must go to zero as $N_f^* - N_f$ goes to zero,
since inside the conformal window it is massless. Therefore if $N_f^* - N_f$ is non-zero but small, the mass of the scalar will be small as well. However, this
argument, based on restoration of conformal symmetry, applies equally well to all massive states, like
the vector meson discussed above. 
All massive states
become massless as $N_f^* - N_f$ goes to zero but we have no information on the ratios. Depending on the rate at which
the masses go to zero, the ratios may stay constant, may go to zero or may go to infinity. Hence there is no a priori
reason for the scalar to be light relative to for example the vector meson even if $N_f^* - N_f$ is small.

\section{Gauge theories inside and outside the conformal window}
\label{conformaland}

The first goal of any lattice simulation of a given model is to determine whether chiral symmetry is spontaneously
broken or not. There are many phenomena that are
markedly different in the two cases and 
a pedagogical overview of the basic differences is given in this section.

The phenomenological motivation limits our interest to conformal gauge theories where a suitably defined
$\beta$-function is not identically zero, but rather has an isolated zero of first order. Hence the prototypical example
of ${\cal N} = 4$ SUSY Yang-Mills theory with an identically zero $\beta$-function is outside the scope of our
discussion. The main difference between an identically zero $\beta$-function and one with an isolated zero is
that in the former case a theory can be constructed at any value of the coupling such that correlation functions fall
off as power-laws on all scales whereas in the latter case there is a single value of the coupling where this is
possible.

\subsection{Infinite volume, zero mass}

The behavior of a spontaneously broken or QCD-like gauge theory at short distances can be described by perturbation theory.
A dynamical scale $\Lambda$ is generated and correlation functions behave as in free theories with logarithmic corrections,
\bea
\langle {\cal O}(x) {\cal O}(0)\rangle = \frac{1}{ x^{2p} } \left( \frac{A}{\log^{2\alpha}(x\Lambda)} + \ldots \right)\;, \qquad \qquad \qquad |x| \ll \Lambda^{-1}
\label{oo}
\eea
with some constants, $A$, $\alpha$ and where $p$ is the engineering or naive dimension of the operator ${\cal O}$. The
constant $\alpha$ is zero if the anomalous dimension of $\cal O$ is zero, for instance if it is a conserved current. In writing eq.~\eqref{oo}
we assume that operators are already renormalized in a suitable scheme at scale $\mu \sim \Lambda$.

The particle spectrum consists of the massless Goldstone bosons originating from the spontaneous breaking of
chiral symmetry as well as a tower of massive bound states. 
The mass of the non-Goldstone bound states are all proportional to $\Lambda$.
Consequently, deep in the large distance regime, more precisely for $\Lambda^{-1} \ll |x|$ 
only power-laws originating from the pions survive. In this regime interaction between the pions can also be neglected and
all correlation functions take on the form of a free theory of pions. This deep infrared limit can formally be realized
by $\Lambda \to \infty$, explicitly taking the mass of all massive bound states to infinity hence decoupling them from the
low lying spectrum of massless (non-interacting) pions. In this sense chirally broken gauge theories are infrared free.
Note however that the weakly interacting degrees of freedom at short distances (gluons and fermions) are different from
the weakly interacting degrees of freedom at large distances (pions).

\begin{figure}
\begin{center}
\includegraphics[width=10cm]{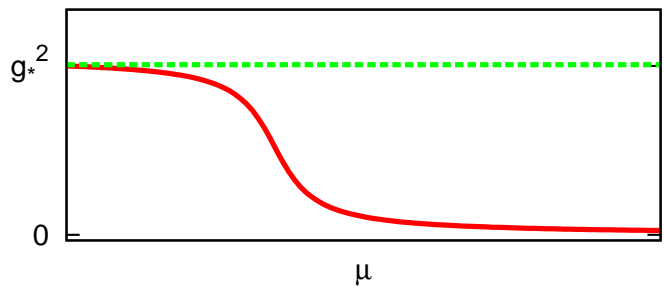}
\end{center}
\caption{Two realizations of the running coupling inside the conformal window. The Lagrangian is the same in the two cases.
The n-point functions fall off as power-laws on all scales (green) or fall off as power-laws for large
distances but their behavior for short distances is described by asymptotic freedom (red). In order to make the
difference clear we will refer to the former (green) as {\em conformal} and the latter (red) as {\em IR-conformal}.}
\label{cft}
\end{figure}

A gauge theory inside the conformal window, 
on the other hand, may behave in one of two distinct ways, see figure \ref{cft}. Note that
the Lagrangian is the same in the two cases.
A suitably defined renormalized running coupling may be constant on all scales, or may reach the fixed point for large
distances only. We will call the former case {\em conformal} and the latter {\em IR-conformal} for definiteness.
For a detailed discussion on the running coupling and its behavior both inside and outside 
the conformal window see section \ref{sec:running}.

In the IR-conformal case a dynamically generated scale $\Lambda$ is present and correlation functions at
short distances behave similarly to a chirally broken theory given by (\ref{oo}). At large distances correlation functions
behave as power-laws,
\bea
\label{ir1}
\langle {\cal O}(x) {\cal O}(0)\rangle = \frac{A}{x^{2p} (x\Lambda)^{2\gamma}} + \ldots\;,\qquad \qquad \qquad |x| \gg \Lambda^{-1}\;,
\eea
where again $p$ is the engineering or naive dimension and $\gamma$ is the anomalous dimension of the operator 
${\cal O}$. 

Clearly, in (\ref{ir1}) one may rescale the coordinate $x$ and operator ${\cal O}$ by
$\Lambda$ to get rid of the dynamical scale at large distances. Hence if,
\bea
\label{resc}
z &=& x\Lambda \nn \\
{\cal O}_{IR}(z) &=& \frac{{\cal O}(z/\Lambda)}{\Lambda^p}
\eea
then in the infrared 2-point functions are simply,
\bea
\label{ir2}
\langle {\cal O}_{IR}(z) {\cal O}_{IR}(0)\rangle = \frac{A}{z^{2p+2\gamma}} + \ldots \;, \qquad \qquad \qquad z \gg 1 \;.
\eea
In the above equation everything is expressed in dimensionless quantities and the dynamical scale $\Lambda$ indeed
dropped out. 

In the second realization of a gauge theory inside the conformal window, 
where correlation functions are power-laws on all scales an
arbitrary dimensionful scale $\Lambda$ may nevertheless be introduced from dimensional analysis of the classical theory.
Then in this case correlation functions behave as equations (\ref{ir1}) and (\ref{ir2}) without corrections
represented by $\ldots$, i.e. for all $x$ and $z$.

One may
imagine regularizing a gauge theory inside the conformal window by a UV-cutoff $\Lambda_{UV}$ or $a^{-1}$ in 
which case all quantities can be measured from the
start in $\Lambda_{UV}$ or $a^{-1}$ units and one would automatically end up with dimensionless quantities. This slight difference
in computation, keeping the dynamical scale $\Lambda$ and only getting rid of it in the infrared by rescaling, or working with
dimensionless quantities from the start is clearly irrelevant as far as the infrared behavior is concerned, but in order
to distinguish the conformal and IR-conformal
scenarios depicted in figure \ref{cft} the dynamical scale $\Lambda$ needs to be kept.

In any case the lack of exponentially falling correlation functions at large distances indicates that all
channels are massless. Note that there is a smooth limit between the two realizations inside the conformal window 
by formally taking $\Lambda \to \infty$, i.e. $\Lambda|x| \to \infty$ while $|x|$ is fixed. 
This limit will turn all correlation functions into power-laws on 
all scales. Even though the lack of a dimensionful scale will of course not make it possible
to measure absolute distance scales, measuring distances relative to each other is still meaningful.
The $\Lambda \to \infty$ limit, as defined here, inside the conformal window simply extends the power-law IR behavior
to all scales but does not alter the (un)particle \cite{Georgi:2007ek} 
content. On the other hand, in a chirally broken gauge theory, this limit 
corresponds to removing all massive states and ending up with only massless pions, i.e. it reduces the number of 
particle species.

\subsection{Finite volume, non-zero mass}

The previous discussion was valid in infinite volume and zero fermion mass. A finite volume and non-zero fermion mass
are both useful tools in lattice calculations as well as unwanted effects that make the distinction between a gauge
theory inside and outside the conformal window
more blurred. The chief reason is that massive fermions introduce massive particle states
and exponentially falling correlation functions even inside the conformal window and finite volume limits the direct
ability to probe the system at large distances.

Nevertheless a finite volume and fermion mass can indeed be used as useful tools since the behavior of a gauge theory
inside or outside the conformal window
differ markedly in well defined regimes. First let us discuss the still massless but finite
volume setup, i.e. the theory is formulated on $T^3 \times R$ with a linear size $L$ for the spatial volume. One
naturally has to impose boundary conditions for both the gauge fields and fermions in the spatial directions and it is
expected that in small volumes, $L\Lambda \ll 1$, the boundary conditions are relevant and may alter the behavior of the
theory substantially whereas for large volumes, $L\Lambda \gg 1$, their influence is expected to be small (either
algebraic or exponential, depending on the quantity in question).

Asymptotic freedom ensures that at
small volume, $L\Lambda \ll 1$, perturbation theory is applicable. In this regime, often called ``femto-world'',
chirally broken and IR-conformal theories 
behave very similarly.
A perturbative Hamiltonian framework can be set up in a straightforward manner
and in this case
all eigenvalues of the Hamiltonian and hence all masses behave as
\bea
\label{both}
M(L) = \frac{1}{L}\left( A + \frac{B}{\log^{2\alpha}(L\Lambda)} + \ldots \right)\; \qquad \qquad \qquad L \ll \Lambda^{-1}\;,
\eea
where the constants $A, B$ and $\alpha$ depend on the quantum numbers of the state and on the boundary conditions.
If the boundary conditions are chosen such that the vacuum is degenerate, tunnelling events will produce splittings
which are small relative to the logarithmic corrections above but are nevertheless reliably calculable for small volume
\cite{Luscher:1982ma, vanBaal:1986ag}.

For large volumes, on the other hand, masses inside and outside the conformal window behave very differently. In
the IR-conformal case we have,
\bea
M(L) = \frac{1}{L}\left( A + \frac{B}{(L\Lambda)^\omega} + \ldots \right)\; \qquad \qquad \qquad L \gg \Lambda^{-1}\;,
\eea
where the exponent $\omega$ may be obtained from the $\beta$-function of the theory, see  section \ref{sec:running}.

On the other hand, if the theory is chirally broken 
the large volume spectrum, $L \gg \Lambda^{-1}$,  will behave markedly differently. In this regime,
familiar as the $\delta$-regime of chiral perturbation theory \cite{Leutwyler:1987ak}, 
there are modes whose volume dependence is
\bea
M(L) = \frac{1}{L(L\Lambda)^2}\left(A + \frac{B}{(L\Lambda)^2} + \ldots \right)\; \qquad \qquad \qquad L \gg \Lambda^{-1}\;
\eea
which will ultimately become the pions at infinite volume and there are also modes whose volume dependence is rather
\bea
M(L) = \Lambda \left( A + \frac{B}{(L\Lambda)^2} + \ldots \right)\;, \qquad \qquad \qquad L \gg \Lambda^{-1}\;
\eea
which at infinite volume become the tower of massive bound states.

Now let us turn to the situation of infinite volume, but finite (bare) fermion mass, $m$. In this case particle states will be
massive even in the conformal case and correlation functions will have an exponential fall off for large distances. 
The masses of gauge singlet particles are of course physical quantities and as such are renormalization group invariant,
however the fermion mass $m$ is not. Let us choose a renormalization scheme for the fermion mass and denote
by ${\tilde m}(m)$ an RG invariant mass. Then the physical masses of particles states will behave as
\bea
\label{conf}
M(m) = A \Lambda \left( \frac{{\tilde m}}{\Lambda} \right)^{\frac{1}{1+\gamma}} + \ldots
\eea
for ${\tilde m}/\Lambda \ll 1$ in conformal theories with $\gamma$ the mass anomalous 
dimension \cite{DelDebbio:2010ze,DelDebbio:2010jy}. The coefficient $A$ 
as well as the function ${\tilde m}(m)$ depends on the renormalization scheme but the exponent $\gamma$ does not.

In the chirally broken case the fermion mass dependence of the Goldstone bosons is determined by the 
$p$-regime of chiral perturbation theory \cite{Gasser:1983yg},
\bea
\label{pi}
M(m) = \Lambda \left( \frac{{\tilde m}}{\Lambda} \right)^{1/2} \left( A + B \frac{{\tilde m}}{\Lambda} + C \frac{{\tilde
m}}{\Lambda} \log \frac{{\tilde m}}{\Lambda} + \ldots \right) 
\eea
and the fermion mass dependence of all other states is
\bea
M(m) = \Lambda \left( A + B \left(\frac{{\tilde m}}{\Lambda}\right)^\alpha + \ldots \right)
\eea
with some exponent $\alpha > 0$, typically $\alpha = 1$. 
It should be noted that the above expressions receive next to leading order
corrections in the chiral expansion which can only be assumed to be small if indeed ${\tilde m}/\Lambda$ is sufficiently
small. Furthermore, at finite ${\tilde m}/\Lambda$ ratio, or in other words at finite Goldstone mass a further
assumption needs to hold, namely that all states are sufficiently heavier than the Goldstone itself. This is because the
conventional chiral Lagrangian from which (\ref{pi}) and expansions of all other low energy quantities are obtained is
only sensitive to the Goldstones as all further states are assumed to be integrated out. 
However at finite fermion mass it may happen that the mass of further states, 
which are non-zero in the chiral limit, become comparable to the mass of the Goldstones in which case they must be
included as correction terms in the chiral Lagrangian. A potential example is the $0^{++}$ meson. Close to the conformal
window direct lattice calculations seem to indicate that indeed the scalar meson does not separate from the Goldstones even at
the smallest fermion masses accessible to numerical simulations.

Apart from expressions like (\ref{pi}) chiral perturbation theory in the p-regime predicts 
relationships between a host of quantities, like the GMOR relation, as well as the fermion mass dependence of decay
constants. In particular the chiral Lagrangian dictates that the decay constant of the Goldstone bosons 
in the chirally broken case behaves as, at leading order,
\bea
\label{fpi}
F(m) = \Lambda \left( A + B \frac{{\tilde m}}{\Lambda} + C \frac{{\tilde m}}{\Lambda} \log \frac{{\tilde m}}{\Lambda} +
\ldots \right)
\eea
where the $A,B,C$ parameters are different from the similarly named parameters in (\ref{pi}), but chiral perturbation
theory establishes relationships between them. In the conformal case, on the other hand,
\bea
F(m) = A \Lambda \left( \frac{{\tilde m}}{\Lambda} \right)^{\frac{1}{1+\gamma}} + \ldots
\eea
is expected for small enough fermion mass $m$.

\section{Mass spectrum as a probe for IR-conformality}
\label{latticeaspects}

So far our discussion was in the continuum. Any lattice simulation is naturally set up in finite 4-volume and finite
lattice spacing. As far as the study of the spectrum is concerned in large volumes
the fermion mass also needs to be finite for technical reasons.
The first goal of any lattice simulation is to establish whether
the simulated theory is inside or outside the conformal window 
at infinite volume and zero fermion mass. This is a non-trivial task
since in order to make use of the continuum expressions which clearly distinguish the two cases, 
one needs to ensure that both the asymptotic requirements for their
validity hold and also that the lattice spacing, $a$, is sufficiently small.
In practice this means that
$\Lambda L \gg 1$ and $M(m)L \gg 1$ for the smallest mass $M(m)$ is required in order to have small finite volume
effects. Furthermore $a\Lambda \ll 1$ and $aM(m) \ll 1$ needs to hold for small cut-off effects.

\subsection{Finite volume effects}

The most direct way to probe the infrared of a given theory on the lattice is to study its mass spectrum in large
volumes keeping the
necessary inequalities as well as one can, given the practical constraints of the available computer. Even though this
approach is theoretically sound the inequalities are hard to fulfill as one approaches the conformal window from
below, as finite volume effects become more and more severe. In practice this means that even though the general rule of
thumb in QCD, $M_\pi(m) L > 4$, ensures small finite volume effects in spectral quantities, in theories close to the conformal
window $M_\pi(m) L > 5$ or even $M_\pi(m) L > 10$ is required \cite{Fodor:2012ty, Amato:2015dqp}.
In addition if one wants to employ infinite volume chiral perturbation
theory, for example (\ref{pi}) or (\ref{fpi}), then $F_\pi L \geq 1$ is also needed which condition is analogous to the general
$\Lambda L \geq 1$ expression. Note that the latter constraint is particularly hard to maintain close to the conformal
window with a small fermion mass because $F_\pi(m)$ varies rapidly as a function of $m$. The coefficient $B$
is apparently larger just below the conformal window than in QCD in equation (\ref{fpi}). 

For a model inside the conformal
window finite volume effects are even more severe and $M_\pi(m) L > 15$ was reported to be necessary to have negligible
finite volume effects at finite fermion mass for the $SU(2)$ model with $N_f = 2$ adjoint fermions \cite{DelDebbio:2015byq}.

Not completely controlling finite volume effects, i.e. having not sufficiently large volumes in the simulations
is not only problematic for applying infinite volume chiral perturbation theory or hyperscaling formulae but also more
generally. We have seen in the previous section that at small $\Lambda L$ IR-conformal and chirally broken theories 
behave very similarly, simply because both are asymptotically free and at not
sufficiently large $\Lambda L$ the simulation can not probe deeply enough in the infrared to distinguish them. The
above mentioned general observation that $F_\pi(m)$ drops more steeply as a function of $m$ for small $m$ 
if the model is closer to the conformal window results in the need for ever larger lattice volumes.

In intermediate volumes, where $\Lambda L \sim 1$ there are no theoretical expectations
for the volume dependence or the fermion mass dependence. Increasing the fermion mass in order to increase $F_\pi(m)$
will ensure $F_\pi(m)L \gg 1$ however $M_\pi(m)$ also grows and the asymptotic expressions for small mass will lose
their validity both inside and outside the conformal window. As a result simulations with practical constraints on the
lattice volume given by the available computer often find themselves between a rock and a hard place: either
intermediate volume or intermediate fermion mass, neither of which has a theoretically sound description.

\subsection{Finite lattice spacing effects}

Furthermore, even though the physical volume in a lattice calculation can be increased at fixed lattice volume by
increasing the lattice spacing via increasing the bare gauge coupling, this will introduce larger cut-off effects and
the $a\Lambda \ll 1$ constraint will hold to a lesser degree. Consequently the conclusions will be less indicative of
the continuum theory and perhaps will be specific to the chosen discretization only. In addition there might be bulk
phase transitions at some critical bare gauge coupling, which is specific to the given discretization and has nothing to
do with the continuum dynamics of the model. In order to draw
conclusions which have a chance to describe the continuum theory the bare coupling $g_0^2$ needs to be smaller than the critical
value and this alone might force the simulation into a regime where the physical volume is not large enough, unless very
large lattice volumes are used which might not be affordable on a given computer.

\subsection{Low lying scalar and chiral perturbation theory}

A further issue, as mentioned, is that if the scalar meson becomes lighter and
lighter, the chiral expansion becomes more and more invalid. Just below the conformal window the scalar meson mass 
seems to become light indeed. In practice it becomes hard to simulate at light enough masses, such that the pion becomes
lighter than the scalar, and this complicates the application of chiral
perturbation theory formulae \cite{Caprini:2005zr, Contino:2010mh, Soto:2011ap, Golterman:2016lsd}.
On the other hand, just inside the conformal window one may need to use very small
fermion masses in order to fit the data with the leading expression (\ref{conf}) and in practice one is forced to use
subleading terms in the fits increasing the number of fit parameters. Similarly, the number of fit parameters will grow
due to cut-off effects as well, in a chirally broken theory the chiral expansion will have new terms which
are vanishing in the continuum but can be sizable at finite cut-off.

\subsection{Selected lattice results}

Since simulations of the mass spectrum close to the conformal window are plagued by the above difficulties, it is all
the more important to gather as much evidence as possible, before conclusions are drawn from numerical data. For instance,
if for a model chiral symmetry breaking appears to take place it is important to verify
this from as many observables as possible. Good chiral fits of the Goldstone mass and decay constant is preferrably
complemented by a verification of the GMOR relation and by checking the Random Matrix Theory predictions for the low lying
Dirac eigenvalues in the $\varepsilon$-regime. Furthermore there are relations between the various chiral fits in the
$p$-regime since the same low energy constants appear in all of them, allowing for powerful consistency checks.  
Similarly, it is desirable to complement the conformal scaling tests of the 
mass spectrum by calculations of the running coupling showing an infrared fixed point (see section \ref{sec:running}) in the
conformal case. 
Also, the
mass anomalous dimension $\gamma$ from the spectrum should be independent from the channel from which it is extracted.
Furthermore it ought to agree with the running mass anomalous dimension at the infrared
fixed point, as well as with the one obtained from the scaling of the Dirac spectrum, 
providing powerful checks in the conformal case too. Note that the study of the Dirac spectrum
has its own source of systematic effects, namely definitive conclusions can only be drawn from small eigenvalues as far
as the infrared is concerned and this range is particularly distorted by finite volume effects \cite{Keegan:2015cba}.

Despite the above complications, the mass spectra of numerous models were calculated on the lattice keeping 
the needed inequalities to varying degrees. 

As far as $SU(2)$ is concerned there is broad agreement that the $N_f = 2$
model in the adjoint representation is conformal, the mass spectrum in particular was studied in detail 
\cite{0812.1467, 0907.3896, 1004.3197, 1004.3206, 1104.4301, 1201.6262, DelDebbio:2015byq}. 
The $N_f = 1$ case was also investigated
\cite{Athenodorou:2014eua} and asymptotic freedom is lost at $N_f
= 2.75$. In the fundamental representation asymptotic freedom is lost at $N_f = 11$. Detailed studies of the particle 
spectrum for $N_f = 2, 4, 6$ are available \cite{Hietanen:2014xca, Amato:2015dqp,Arthur:2016dir} 
with $N_f = 6$ being thought to be at around the lower
end of the conformal window. Severe finite volume effects at $N_f = 6$ however prohibited a conclusive result as to
whether the model is chirally broken or already inside the conformal window. 

The gauge group $SU(3)$ was studied on the lattice by many groups. Since the fundamental representation is particularly
familiar from QCD applications, this model was the first to be investigated in detail. The $N_f = 6$ model is certainly
outside the conformal window. The mass spectrum of the $N_f = 8$ model was studied extensively
\cite{Fodor:2009wk, Aoki:2013xza, Aoki:2014oha, Rinaldi:2015axa, Appelquist:2016viq, Appelquist:2014zsa}, results for
both $N_f = 9$ \cite{Fodor:2009wk} and $N_f = 10$ \cite{Appelquist:2012nz} are available as well as $N_f = 12$
\cite{Fodor:2011tu, Aoki:2012eq, Aoki:2013zsa}. 
There seems to be disagreement about the $N_f = 12$ model, whether it is already inside or just
below the conformal window and the study of the running coupling does not seem to resolve this issue (see section
\ref{sec:running}).
Beyond the fundamental representation the
most promising candidate model from a phenomenological point of view is the sextet with $N_f = 2$ flavors
\cite{Sannino:2004qp,Hong:2004td,Dietrich:2005jn}. The mass
spectrum was investigated in detail \cite{Fodor:2012ty,Fodor:2014pqa, Fodor:2015eea, Fodor:2015vwa, Drach:2015sua}, 
along with various chiral properties.
The results seem to be consistent with chiral symmetry breaking although
see also \cite{Drach:2015sua}.

The adjoint of $SU(2)$ or the sextet of $SU(3)$ are the two index symmetric representations and generalizing it further,
a first study of $SU(4)$ gauge theory with $N_f = 2$ flavors in the two index symmetric was recently performed
\cite{DeGrand:2015lna}.

As mentioned in section \ref{strong} one of the most important conclusions drawn from lattice studies of gauge
theories close to the conformal window is the appearance of a light composite scalar meson. Here by light we mean
its mass $m_\sigma$ relative to the mass $m_\varrho$ of the vector meson. In the $SU(3)$ model with
$N_f = 8$ fundamental fermions approximately $m_\sigma / m_\varrho \sim 1/2$ was observed, 
whereas with $N_f = 2$ sextet fermions approximately $m_\sigma
/ m_\varrho \sim 1/4$. These observations make it plausible that a composite Higgs may emerge from a near-conformal gauge
theory with its $125\; GeV$ mass obtained after electro-weak corrections are taken into account \cite{Foadi:2012bb}.

Beyond the unitary gauge group, the mass spectrum of $SO(4)$ was studied \cite{1211.5021} with 
$N_f = 2$ flavors in the fundamental representation, showing consistency with chiral symmetry breaking.

Due to the practical difficulties alternative approaches were also explored in lattice calculations. One area
where lot of effort was concentrated is the calculation of the $\beta$-function of the models, outlined in the next
section.

\section{Running coupling as a probe for IR-conformality}
\label{sec:running}

The basic idea behind the running coupling studies is that an IR fixed point would be characterized by the property that the running coupling goes to a finite value in the limit of zero energy. Typically a very general definition of running coupling is adopted: any observable $g^2(\mu)$ which depends on a single energy scale $\mu$ and which admits the following perturbative expansion
\begin{gather}
g^2(\mu) = g_{\RS{}}^2(\mu) + \sum_{n=1}^\infty c_n g_{\RS{}}^{2n}(\mu) \ ,
\label{eq:running:general_expansion}
\end{gather}
valid for $\mu \to \infty$ is said to be a running coupling. A reference renormalization scheme $\RS{}$ has to be assumed in this definition. The $\overline{MS}$ can be considered for definiteness, but other schemes might be used as well. It is worth reminding that the above series is only formal, it does not converge and it does not imply analyticity. We will say that a given coupling $g^2(\mu)$ is a \textit{good probe for IR-conformality} if it diverges in the $\mu \to 0$ limit in theories with spontaneous chiral symmetry breaking (S$\chi$SB) and goes to a finite nonzero value in IR-conformal theories. Throughout this section we will assume that we are setting the quark masses equal to zero, and IR-conformality is possibly broken only by a finite volume.

Unfortunately eq.~\eqref{eq:running:general_expansion} is not enough to guarantee that the running coupling $g^2(\mu)$
is a good probe for IR-conformality. It is easy to construct observables $g^2(\mu)$ satisfying~\eqref{eq:running:general_expansion} that \textit{do not} diverge in theories with S$\chi$SB. In general, given a running coupling $g^2(\mu)$ that diverges in the $\mu \to 0$ limit, it is always possible to construct another running coupling
\begin{gather}
   \tilde{g}^2(\mu) = \frac{g^2(\mu)}{1+g^2(\mu)}
   \label{eq:running:alg1}
\end{gather}
that goes to 1 in the $\mu \to 0$ limit. Later on we will show how a coupling defined in terms of the vector-current two-point function does not diverge in the IR limit even if chiral symmetry breaks, exactly because of pion physics.

It is also easy to produce examples of couplings that diverge in the IR limit in case of IR-conformality. Let us assume that $g^2(\mu)$ behaves in the IR limit accordingly to standard Wilsonian RG behaviour
\begin{gather}
   g^2(\mu) \simeq g_*^2 - A_\omega (\mu/\Lambda)^{\omega} + \dots \ ,
\end{gather}
where $A_\omega$ and $\omega$ are positive numbers. In particular $\omega$ is related to the anomalous dimension of the first irrelevant operator at the IR fixed point. Define now the following coupling
\begin{gather}
   \tilde{g}^2(\mu) = - \frac{b_0 g^6(\mu)}{\mu \frac{\partial g^2(\mu)}{\partial \mu}} \ ,
   \label{eq:running:alg2}
\end{gather}
where $-b_0$ is the first coefficient of the expansion of the beta function around $g^2=0$ and ensures the validity of the representation~\eqref{eq:running:general_expansion}. In the IR limit
\begin{gather}
   \tilde{g}^2(\mu) \simeq \frac{b_0 g^6_*}{\omega A_\omega} (\mu/\Lambda)^{-\omega} + \dots \ ,
\end{gather}
which shows that $\tilde{g}^2(\mu)$ diverges. This example shows that the standard Wilsonian RG treatment does not work for the coupling $\tilde{g}^2(\mu)$. The reason is that Wilsonian RG assumes regularity properties that might not hold, and in fact one should not expect to be valid especially if the considered theory is strongly coupled. Typically couplings defined at high energies and satisfying eq.~\eqref{eq:running:general_expansion} capture the interaction strength between quarks, and they have nothing to do with large distance physics. Both in the case of spontaneous $\chi$SB and IR-conformality the large-distance degrees of freedom are in fact colorless and can be approximated as quark bound states in case of a weakly coupled Banks-Zaks fixed point.\cite{Caswell:1974gg,Banks:1981nn}

We believe that whenever a coupling $g^2(\mu)$ satisfying eq.~\eqref{eq:running:general_expansion} is proposed to study IR-conformality, then a proof of the property that $g^2(\mu)$ is also a good probe for IR-conformality should be provided which is not based merely on perturbative Wilsonian RG, but maybe on more general effective-theory analysis, before definitive conclusions are drawn. Surprisingly enough this logical issue has been largely ignored in the literature. We will review some possible definitions of running couplings, trying to highlight what we know or we do not know about their IR-behaviour.

\subsection{Static potential}
\label{subsec:running:static}

The force $F(r)$ between static quarks can be defined in terms of rectangular Wilson loops with size $r \times t$ as
\begin{gather}
F(r) = \lim_{t \to \infty} \frac{1}{t} \frac{\partial}{\partial r} \ln W(r,t) \ .
\end{gather}
We assume for simplicity that we have already taken the zero-mass and infinite-volume limits. At small $r$ the force between static quarks has a perturbative expansion
\begin{gather}
F(r) = - \frac{k \, g_{\RS{}}^2(r^{-1}) + O(g_{\RS{}}^4)}{r^2} \ ,
\end{gather}
where $k$ is a positive constant. A running coupling can be defined as
\begin{gather}
g^2_F(\mu) = - \left. k^{-1} r^2 F(r) \right|_{r=\mu^{-1}} \ .
\end{gather}
The static force provides a physically motivated definition of the running coupling, at least for short distances or in other words in the perturbative regime. If the model exhibits S$\chi$SB, the force is governed by the dynamics of the effective string at intermediate distances and $F(r) \simeq - \sigma$. At large enough distances, in theories that generate string-breaking (like QCD), the effective string is broken by generation of a light quark-antiquark pair, and each dynamical quark binds to a static one forming heavy-light mesons. In this regime $F(r)$ becomes the force between these mesons, rather than between static quarks. At asymptotically large distances it is dominated by one-pion exchange. Since we are in the chiral limit, the pion is massless and the induced interaction is Coulombic, i.e. the force vanishes proportionally to $r^{-2}$. Therefore the coupling $g^2_F(\mu)$ grows quadratically at intermediate distances and goes to a constant at very large distances. It is worth mentioning that this problem is avoided in theories with a residual center symmetry (e.g. confining theories with fermions in the adjoint representation): in this case string breaking does not occur and the running coupling grows quadratically at asymptotically large distances.

In case of IR conformality, the force is expected to be Coulombic at large distance and the coupling $g^2_F(\mu)$ is expected to go to a non-zero finite value. In conclusion, even though in some intermediate regime the quantity $g^2_F(\mu)$ is expected to behave differently in case of IR-conformality and S$\chi$SB, its behavior at asymptotically large distance is not sufficient to unambiguously differentiate between the two cases. Empirically one sees that the regime in which the effective string breaks is very hard to reach in typical numerical simulations, and in practice only short and intermediate distances are explored. Earlier results using variations of this scheme include e.g. Creutz ratios\cite{Fodor:2009wk}, or the twisted Polyakov loop (TPL) coupling\cite{Lin:2012iw,Itou:2012qn} to investigate IR-conformality. It is instructive to notice that the TPL coupling is expected to go to a constant in the low-energy limit even in pure Yang-Mills theory\cite{deDivitiis:1994yz}, because of an algebraic cancelation very similar in spirit to the one in eq.~\eqref{eq:running:alg1}. In the case with dynamical fermions a similar saturation effect is expected\cite{Lin:2012iw,Itou:2012qn}.

\subsection{Vector current}
\label{subsec:running:vector}

We consider the two-point function of the non-singlet vector current, calculated in infinite volume:
\begin{gather}
C_V(x) = \langle V_\mu^a(x) V_\mu^a(0) \rangle \ , \qquad V_\mu^a(x) = \bar{\psi} \tau^a \gamma_\mu \psi(x) \ .
\end{gather}
At small $x$ the two-point function admits a perturbative expansion $x^6 C_V(x) = c_0 + c_1 g_{\RS{}}^2(x^{-1}) + \dots$ where the $c_0$ and $c_1$ coefficients can be analytically worked out (see section \ref{conformaland}). Therefore one can define a legitimate running coupling as follows
\begin{gather}
g_{V}^2(\mu) = \left. \frac{x^6 C_V(x) - c_0}{c_1} \right|_{x=\mu^{-1}} \ .
\label{eq:running:gtildeV}
\end{gather}
This running coupling has never been used in studies of the conformal window. However it possesses very interesting
features that are worth highlighting. If the theory is IR-conformal, the large distance behaviour is determined by the
scaling dimension of the vector current. Since $V_\mu^a(x)$ is a conserved current, its scaling dimension is equal to
its engineering one. This means that the vector two-point function decays like $x^{-6}$ at large distances. Therefore
the coupling $g_{V}^2(\mu)$ goes to a constant in the $\mu \to 0$ limit as expected. If chiral symmetry is spontaneously
broken, then the vector current couples to two-pion states at large distance. If $\pi$ is the pion field, at the leading
order in chiral perturbation theory, the vector current is represented by the operator $\tr \tau^a \pi \partial_\mu \pi$
up to total derivatives.\cite{Gasser:1983yg} It is easy to check by power counting that the vector two-point function
decays like $x^{-6}$ (one $x^{-2}$ per pion propagator and one $x^{-1}$ per derivative). Therefore the running coupling
$g_{V}^2(\mu)$ goes to a constant in the $\mu \to 0$ limit even if chiral symmetry is spontaneously broken. Notice that
this constant is predicted by chiral perturbation theory.

\subsection{Schr\"odinger functional (SF) coupling}
\label{subsec:running:SF}

Most studies which aim at determining IR conformality in gauge theories have used finite-volume renormalization schemes.
The idea is to define the running coupling as some observable calculated in a hypercubic box and to identify the
renormalization scale $\mu$ with the inverse of the box size $L$. This approach has the advantage to remove or
dramatically reduce two sources of systematic errors in typical lattice simulations: \textit{(1)} the infinite-volume
extrapolation, and \textit{(2)} the chiral extrapolation. In finite volume, if boundary conditions are properly chosen,
the Dirac operator has a gap even in the massless limit and simulations at the chiral point are possible. If fermions
with a residual chiral symmetry are employed then one can simulate exactly at zero bare mass. In case of Wilson fermions
the chiral limit is reached at an unknown value of the bare mass which can be found by interpolation (rather than
extrapolation). In these kinds of calculations one still has systematic errors that come from the continuum extrapolation, on which we will comment later. It is worth noticing that in order to ensure a perturbative expansion of the type~\eqref{eq:running:general_expansion} one needs to use boundary conditions such that the vacuum is unique at tree level. One can relax this condition by choosing boundary conditions such that the vacuum is degenerate at tree-level but the degeneracy is completely lifted at one-loop, provided that more general expansions than~\eqref{eq:running:general_expansion} are considered.\cite{GonzalezArroyo:1981vw}

One can consider a hypercubic box with periodic boundary conditions in the three spatial directions, and SF boundary conditions\cite{Rossi:1979jf,Rossi:1982af} for the gauge field at the boundaries $x_0=0$ and $x_0=L$. Typically one chooses
\begin{gather}
   A_k(0,\vec{x}) = \frac{\eta \lambda_1}{L} \ , \quad A_k(L,\vec{x}) = \frac{\lambda_0 - \eta \lambda_1}{L} \ ,
\end{gather}
where $\lambda_0$ and $\lambda_1$ are color matrices and $\eta$ is a free parameter. Also the fermion fields satisfy some appropriate boundary conditions, whose explicit form plays no role in the present discussion. The boundary conditions induce a background chromomagnetic field. If the background field is properly chosen, uniqueness of the tree-level vacuum is ensured. The variation of the free energy with respect to the boundary fields turns out to be proportional to the inverse of the squared coupling, and can be used to define a running coupling\cite{Luscher:1993gh}, as in
\begin{gather}
   \left. \frac{1}{g_{\text{SF}}^2(\mu)} \right|_{\mu = L^{-1}} = k \left. \frac{d}{d \eta} \right|_{\eta=0} \ln Z_{\text{SF}}(\eta) \ ,
   \label{eq:running:SFcoupling}
\end{gather}
where $Z_{\text{SF}}$ is the partition function with SF boundary conditions and $k$ is a constant that ensures the correct normalization. The renormalizability of QFT with SF boundary conditions and the existence of the continuum limit of the SF coupling are nontrivial issues and have been discussed in the literature.\cite{Symanzik:1981wd,Luscher:1985iu,Luscher:1992an,Luscher:1993gh,Kennedy:2015uob}

Empirically one observes that in pure Yang-Mills and QCD the SF coupling diverges at $L \to \infty$. In pure Yang-Mills one can easily argue that this is in fact the case by using the existence of a mass gap.\cite{Heitger:2001hs} In a theory with spontaneous $\chi$SB, the leading contribution to the running coupling at large volume will come from multi-pion exchange between the two boundaries or from pions traveling around the periodic direction. These contributions are powers in $L$, and depending on the exponent they could lead to a vanishing, finite or divergent behaviour of the running coupling at low energies. In principle this power can be determined by representing the SF running coupling in terms of operators of the chiral Lagrangian. It is interesting to notice that this issue has not been addressed from the theoretical point of view.

In case of IR conformality one would like to argue that the SF running coupling must go to a constant in the $L \to \infty$ limit. This is most probably the case, but the issue is far from being completely trivial. By working out the derivative with respect to the boundary conditions in eq.~\eqref{eq:running:SFcoupling} one finds out that the SF running coupling can be represented in terms of expectation values of operators on the boundaries
\begin{gather}
   \left. \frac{1}{g_{\text{SF}}^2(\mu)} \right|_{\mu = L^{-1}} = \frac{k_0}{L} \int_{L^3} d^3x \  \langle \tr \lambda_1 F_{0k}(0,\vec{x}) \rangle_{\text{SF}} + \frac{k_L}{L} \int_{L^3} d^3x \  \langle \tr \lambda_1 F_{0k}(L,\vec{x}) \rangle_{\text{SF}} \ .
   \label{eq:running:SFcoupling2}
\end{gather}
In fact this is the way in which the SF running coupling is calculated in numerical simulations. Notice that the operator $\tr \lambda_1 F_{0k}$ is not gauge invariant, but this is not a problem as the boundary conditions are not invariant under gauge transformations. At the fixed point, the bulk theory is scale invariant. The finite volume breaks scale invariance softly, which means that the trace of the energy momentum tensor is zero in the bulk, but not necessarily on the SF boundary. If no dynamical scale is generated on the boundary, then by dimensional analysis the expectation value of $\tr \lambda_1 F_{0k}$ should be proportional to $L^{-2}$ yielding a finite limit for the running coupling for $L \to \infty$. However notice that the boundary field is not invariant under (3-dimensional) dilations, therefore we expect the trace of the energy momentum tensor to get a non-vanishing contribution at the boundary, and a dynamical scale could be generated if the relevant or marginal operators of the boundary theory get anomalous dimensions. This issue might well turn out to be trivial, but it is surely worth to be analyzed in detail.

In conclusion it looks very plausible that the SF coupling turns out to be a good probe for IR conformality, however more theoretical work is needed in order to understand its low-energy limit. The SF coupling has been widely used to investigate IR conformality in various theories mostly until 2013, and then it has been almost completely replaced by the much more precise gradient-flow coupling. All SF-coupling studies\cite{Hietanen:2009az,Bursa:2009we,DeGrand:2011qd,Rantaharju:2015yva} agree on the existence of an IR fixed point in $SU(2)$ with 2 adjoint fermions. Concerning $SU(2)$ with $N_f$ fundamental fermions, the SF-coupling runs away for $N_f=4$\cite{Karavirta:2011zg}, and an IR-fixed point is found for $N_f=10$\cite{Karavirta:2011zg}. The case $N_f=6$ collects evidence in favour of slow running of the SF-coupling\cite{Bursa:2010xn,Karavirta:2011zg} and against it\cite{Appelquist:2013pqa}. 
The $SU(3)$ gauge theory with 8 fundamental fermions collected evidence for strong running of the SF-coupling\cite{Appelquist:2007hu,Appelquist:2009ty}. The same studies report evidence for an IR-fixed point in the $SU(3)$ gauge theory with 12 fundamental fermions. Slow running of the SF-coupling has been reported also in the $SU(3)$ theory with 2 sextet fermions\cite{Shamir:2008pb,DeGrand:2010na,DeGrand:2012yq}, in the $SU(3)$ theory with 2 adjoint fermions and in the $SU(4)$ theory with 6 antisymmetric two-index fermions\cite{DeGrand:2013uha} and in the $SU(4)$ theory with 2 symmetric two-index fermions\cite{DeGrand:2012qa}.

\subsection{Gradient flow (GF) coupling}
\label{subsec:running:GF}

The gauge field $B_t$ at positive flowtime $t$ is defined as a function of the fundamental gauge field $A$ through the differential equation
\begin{gather}
   \partial_t B_{t,\mu} = D_{t,\mu} G_{t,\mu\nu} \ , \qquad
   B_{0,\mu} = A_\mu \ ,
\end{gather}
where $D_{t,\mu}$ and $G_{t,\mu\nu}$ are respectively the covariant derivative and the field strength tensor built with
the gauge field $B_{t,\mu}$. The GF coupling\cite{Luscher:2010iy,Fodor:2012td} is defined in a finite hypercubic box with some given boundary conditions as
\begin{gather}
g^2_G(\mu) = \mathcal{N}(c) \left. t^2 \langle \tr G_t^2 \rangle \right|_{\mu = L^{-1} = c (8t)^{-1/2}} \ ,
\label{eq:running:GFcoupling}
\end{gather}
where $c$ is some arbitrarily chosen constant and $\mathcal{N}(c)$ gives the correct normalization of the coupling. The boundary conditions are often chosen in such a way that the perturbative expansion is non-degenerate and a representation of the type~\eqref{eq:running:general_expansion} holds, however this is not necessary to define a possible probe for IR conformality. The existence of the continuum limit of the GF coupling is non trivial and we refer to the relevant literature for its proof.\cite{Luscher:2011bx}

As for the SF coupling, no proof is available of the expectation that the GF coupling diverges in case of S$\chi$SB. As
for the SF functional one might want to represent the GF coupling in terms of operators in the framework of chiral
perturbation theory. This might allow us to understand the IR behaviour of the coupling in terms of pion physics. Notice
that operators at some nonzero but fixed flowtime are non-local, but the range of nonlocality is small with respect to
the pion Compton length. Therefore they can be represented as local operators in terms of the pion
fields\cite{Bar:2013ora}. However the IR behavior of the GF coupling is obtained in the $t \to \infty$ limit and it is
not obvious \textit{a priori} that this regime is correctly captured by chiral perturbation theory.

In the case of IR-conformality, one can argue that operators at positive flowtime do not get anomalous dimensions, and therefore $\langle \tr G_t^2 \rangle$ vanishes proportionally to $t^{-2}$ in the large $t$ limit. This immediately implies that the GF coupling goes to a constant in the IR limit. In order to see this it is useful to think of the flowtime as a real coordinate\cite{Luscher:2011bx}. Operators at positive flowtime are mapped into local operators in a 5-dimensional theory with boundary ($t=0$). At the IR fixed point, the original 4-dimensional theory becomes scale invariant. One would like to understand whether the full 5-dimensional theory is scale invariant as well. Notice that the GF equation is scale invariant which implies that the bulk theory is scale invariant. Moreover the bulk theory is classical so no anomalous dimensions will be generated. Because of the interaction of the 4-dimensional theory with the bulk theory, new boundary operators are generated. In order to estabilish scale invariance of the full 5-dimensional theory, one needs to make sure that no relevant operators are generated on the boundary because of the interaction with the bulk. This is surely true if the fixed point is sufficienlty weakly coupled. It would be interesting to understand whether stronger results could be estabilished, e.g. whether the absence of chirally-invariant relevant operators in the original 4-dimensional theory implies the absence of relevant interaction boundary operators.

The GF coupling has the great advantage over other couplings to come with small statistical errors in numerical simulations. For this reason it has practically become the coupling of choice in studies of IR-conformality. Concerning the $SU(3)$ gauge theory with $N_f$ fundamental fermions, clear indication for fast running has been observed for $N_f=4,8$\cite{Cheng:2014jba,Hasenfratz:2014rna,Fodor:2015baa}. Slow running has been confirmed for $N_f=12$\cite{Lin:2015zpa} even though the authors observe no compatibility with IR-conformal finite-size scaling. Compatibility with an IR fixed point has been observed for $SU(2)$ with 2 adjoint fermions\cite{Rantaharju:2015cne}, consistently with previous studies. Studies of the running coupling of $SU(3)$ with two sextet fermions show some tension\cite{Hasenfratz:2015ssa,Fodor:2015zna}. Interestingly the studies of the spectrum of this theory seem to point towards S$\chi$SB with strong non-QCD like features.

\subsection{Nucleon mass}
\label{subsec:running:nu}

Finally we give an example of a possible coupling whose IR behaviour is very easy to predict and is deeply related to the physics that we would like to probe. We consider a generic gauge theory coupled to a number of massless Dirac fermions in some representation of the gauge group, such that twisted boundary conditions \text{\`a la} 't Hooft\cite{tHooft:1979uj} can be used. We consider a $T^3 \times R$ box with linear spatial size equal to $L$, and with twisted boundary conditions in some of the spatial planes. In this setup it is possible to extract the mass gap $M(L)$ in the sector at baryon number equal to one from the long-distance behaviour of some properly defined two-point function. At small volume the mass gap has a perturbative expansion:
\begin{gather}
L M(L) = c_0 + c_1 \, g_{\RS{}}^2(L^{-1}) + O(g_{\RS{}}^4) \ .
\end{gather}
where the $c_0$ and $c_1$ coefficients are calculable analytically. Therefore one can define a running coupling satisfying eq.~\eqref{eq:running:general_expansion} as follows
\begin{gather}
g_{M}^2(\mu) = \left. \frac{ L \, M(L) - c_0 }{c_1} \right|_{L=\mu^{-1}} \ .
\label{eq:running:gtildeM}
\end{gather}
If chiral symmetry is spontaneously broken, then the gap is expected to survive in infinite volume and the coupling diverges. If the theory is IR-conformal, the gap is expected to vanish proportionally to $1/L$, and the coupling goes to a constant.
A similar construction with the pion mass instead of the nucleon mass would provide a running coupling that behaves in a funny way. In fact in the chiral limit the pion mass vanishes in the infinite-volume limit irrespectively of the long distance properties. The chiral symmetry broken and IR-conformal scenarios are discriminated by how fast the pion mass vanishes. In case of IR-conformality the pion mass would vanish like $L^{-1}$ as any other mass. In contrast the large volume limit in the case of spontaneous chiral symmetry breaking (a.k.a. $\delta$ regime) is dominated by the rotor physics and the pion mass vanishes like $L^{-3}$, as already discussed in section~\ref{conformaland}. A running coupling defined like in \eqref{eq:running:gtildeM} with the pion mass would go to a non-zero constant in the case of IR-conformality and would \textit{vanish} in the case of spontaneous chiral symmetry breaking, in the $\mu \to 0$ limit. One might be tempted to use other mesonic states other than the pion, for instance the mass of the ground state in the non-singlet vector channel. However notice that in case of chiral symmetry breaking and in large volume, this state is a state of two weakly-interacting pions (and not the $\rho$ resonance). Therefore its energy vanishes as $L^{-3}$, like for the single-pion state.

\subsection{Continuum limit of finite-volume couplings}

We want to comment on the largest source of systematic error in running coupling determinations, i.e. the continuum extrapolation. The material discussed here is trivial for lattice practitioners, but might be useful for physicists of other communities who try to interpret the meaning and quality of lattice data. We discuss the setup that is usually employed to calculate the running coupling in finite volume schemes. Again, this means that the running coupling is measured in a finite hypercubic box with length $L$ and the renormalization scale is identified with $L^{-1}$. The primary observable that one measures on the lattice is not the running coupling itself, but rather the so-called step-scaling function $\sigma(u,s)$ defined by the implicit equation
\begin{gather}
   \sigma(u,s) = \left. g^2(sL)\right|_{g^2(L)=u} \ ,
   \label{eq:stepscaling}
\end{gather}
in terms of which one can define the discrete beta function e.g. as
\begin{gather}
   B(u,s) = \left. \frac{g^2(sL)-g^2(L)}{\ln s} \right|_{g^2(L)=u} = \frac{\sigma(u,s) - u}{\ln s} \ .
\end{gather}
Note that the $s \to 1$ limit reproduces the infinitesimal beta function familiar from continuum perturbation theory.
However on the lattice a finite and rational value for $s$ is chosen, since the lattice size in units of the lattice
spacing is always an integer. In an asymptotically free theory $B(u,s)$ is always positive. Moreover $B(u,s)$ vanishes
when it approaches fixed points. Often in the lattice literature the ratio $\sigma(u,s)/u$ is considered instead of the discrete beta function. 

In lattice simulations one measures $\Sigma(u,s,N)$, a discretized version of $\sigma(u,s)$, which is function of the lattice size in lattice units $N=L/a$. The continuum limit $a\to 0$ is reached if $N = L/a$ goes to infinity
\begin{gather}
   \hat{g}^2(g_0,N) = u \ ,
   \label{eq:scalesettingwithu}
\end{gather}
determines $g_0$ as a function of the target value $u$ of the running coupling and the number of points $N$. We will denote $g_0(N,u)$ this function. Notice that in the continuum the running coupling $g^2(L)$ is actually a function of $L \Lambda$, where $\Lambda$ is the dynamically generated scale. Assuming that $g^2(L)$ is a monotonous function of $L \Lambda$, fixing the value of the running coupling through eq.~\eqref{eq:scalesettingwithu} actually means to fix the box size $L$ in units of $\Lambda^{-1}$.

In analogy with eq.~\eqref{eq:stepscaling}, the step-scaling function is defined in the lattice discretized theory as
\begin{gather}
   \Sigma(u,s,N) = \left. \frac{\hat{g}^2(g_0,sN)}{u} \right|_{\hat{g}^2(g_0,N) = u} \ .
\end{gather}
The continuum limit $a\to 0$ is reached if $N = L/a$ goes to infinity while $L$ is kept fixed, i.e. while condition~\eqref{eq:scalesettingwithu} is satisfied. This happens at the Gaussian fixed point
\begin{gather}
   \lim_{N \to \infty} g_0(N,u) = 0 \ ,
\end{gather}
which is the fixed point that defines the continuum limit in asymptotically free theories. Therefore the step-scaling function in the continuum is given by
\begin{gather}
   \sigma(u,s) = \lim_{a/L \to 0} \Sigma(u,s,a/L) \ .
\end{gather}
In practice this limit is obtained by fitting the following simple functional form to the data
\begin{gather}
   \Sigma(u,s,a/L) = \sigma(u,s) + \alpha(u,s) \frac{a^2}{L^2} \ .
   \label{eq:continuumfit}
\end{gather}
This is motivated by the Symanzik effective description of the lattice artefacts\cite{SymanzikEffective} in an $O(a)$-improved setup. At very small values of $a$ the physics at the cutoff scale is always governed by the Gaussian fixed point, irrespectivily of the existence of an IR fixed point. The truncation in~\eqref{eq:continuumfit} assumes that $O(a^4)$ terms are subleading. If the theory is IR conformal and $u$ is close enough to its IR fixed-point value for typical values of $a/L$, which means $a \gg \Lambda^{-1}$, it is reasonable to expect that higher orders become important. In fact detailed study of the systematic errors due to the truncation of the series in~\eqref{eq:continuumfit} typically show that the $O(a^4)$ cannot be neglected and this generally results in very large systematic errors for the continuum extrapolation close enough to the IR fixed point. Clearly the value of the coupling at which the validity of the truncation breaks down depends on the particular discretization of the action. However the general message to take home is that even with step-scaling procedure large lattices are still necessary in order to investigate IR-conformality.

\subsection{Anomalous dimension of $\bar{\psi}\psi$}

Finite volume and the step-scaling procedure can be used also to calculate the renormalization factor of $\bar{\psi}\psi$ (or of the mass), from which one can extract the corresponding anomalous dimension $\gamma(g)$ as a function of the running coupling, or its value $\gamma_*$ at the fixed point if the fixed-point value of the coupling is known. This technique has been widely used in a variety of theories which we will not review here in detail.
\cite{Bursa:2009we,DeGrand:2010na,Bursa:2010xn,DeGrand:2011qd,DeGrand:2012qa,DeGrand:2012yq,DeGrand:2013uha,Perez:2015yna}
The $\bar{\psi}\psi$ anomalous dimension can be extracted also from other techniques, which do not require knowledge of
the value of the coupling at the fixed point, including fit of infinite-volume masses to hyperscaling relations, finite-size scaling analysis of masses\cite{DeGrand:2009hu ,DeGrand:2011cu, Fodor:2012et, Cheng:2013xha, Lombardo:2014pda, Athenodorou:2014eua, Lin:2015zpa} and power-law fits of the spectral density of the Dirac operator. \cite{Cheng:2011ic,Patella:2012da,Cheng:2013eu,Athenodorou:2014eua,Fodor:2014zca,Perez:2015yna}

\section{Outlook}
\label{outlook}

Lattice simulations of 4-dimensional non-abelian gauge theories close to the lower edge of the conformal window are
difficult. There are systematic effects which are rather mild for QCD but become dominant as the conformal window is
approached from below. We have reviewed two approaches in detail (1) study of the mass spectrum at finite fermion mass
in infinite volume; and (2) study of the running coupling in the massless case in finite volume. There are
numerous other very useful and promising approaches but the no-free-lunch theorem seems to apply: if one aspect of the
calculation manages to suppress some unwanted systematic effect, another aspect will unavoidably bring back a
different, potentially more severe, one. In order to judge the quality of any given lattice result there is no simple
rule of thumb to apply but rather all the potential sources of systematic effects, specific to the given approach used,
have to be scrutinized. This is, admittedly, not an easy task. Not fully controlling all
systematic effects leads to lattice results which are on occasion not fully consistent, 
but we believe further work in understanding these both theoretically and algorithmically 
will eventually provide a mature set of results similar to QCD.

What has nevertheless consistently emerged from the non-perturbative lattice investigations is important for model
building and phenomenology. The particle
spectrum of models close to the conformal window seems to contain a light scalar 
(relative to for example the vector meson) which might be interpreted as a composite Higgs particle. How the
other composite particles of the spectrum of any potential strongly interacting model 
fit into the Standard Model or extensions thereof
is not entirely clear at the moment. Hopefully further lattice investigations together with progress on the
experimental side will provide further constraints to help separate the viable from the non-viable models.

\section*{Acknowledgments}

This work was supported in part by OTKA under grant OTKA-NF-104034. We would like to thank Maurizio Piai, Luigi Del Debbio and Marina Marinkovic for comments. We also thanks Julius Kuti, Alberto Ramos, Rainer Sommer, Stefan Sint, Martin L\"uscher and David Lin for discussions on some specific issues.

\bibliographystyle{ws-ijmpa}
\bibliography{full}

\end{document}